# Substrate Induced van der Waals Force Effect on the Stability of Violet Phosphorous

Sarabpreet Singh[1], Mahdi Ghafariasl[1], Hsin-Yu Ko[2], Sampath Gamage[1], Robert A. Distasio Jr.[2], Michael Snure[3], Yohannes Abate[1]

[1] Department of Physics and Astronomy, University of Georgia, Athens, Georgia 30602, USA

[2] Department of Chemistry and Chemical Biology, Cornell University, Ithaca, NY 14850, USA

[3] Air Force Research Laboratory, Sensors Directorate, Wright Patterson Air Force Base, Ohio 45433, USA

## ABSTRACT

The van der Waals (vdWs) forces between monolayers has been a unique distinguishing feature of exfoliable materials since the first isolation of graphene. However, the vdWs interaction of exfoliable materials with their substrates and how this interface force influences their interaction with the environment is yet to be well understood. Here, we experimentally and theoretically unravel the role of vdWs forces between the recently rediscovered wide band gap p-type vdW semiconductor violet phosphorus (VP), with various substrates (including, $SiO_2$, mica, Si, Au) and quantify how VP stability in air and its interaction with its surroundings is influenced by the interface force. Using a combination of infrared nanoimaging and theoretical modeling we find the vdWs force at the interface to be a main factor that influences how VP interacts with its surroundings. In addition, the hydrophobicity of the substrate and the substrate surface roughness modify the vdWs force there by influencing VP's stability. Our results could guide in the selection of substrates when vdW materials are prepared and more generally highlight the key role of interface force effects that could significantly alter physical properties of vdWs materials.

# INTRODUCTION

In a visionary speech in 1960, R. Feynman memorably highlighted the significance of non-covalent forces that dominate materials at the nanometer length scale[1]. Among these forces, interactions due to correlated electronic fluctuation, so called van der Waals (vdWs) forces[2-7], have come to be a key distinguishing feature of layered materials that have taken central stage in contemporary condensed matter physics. This weak vdWs force that holds layered materials together makes it possible not only to isolate one atom thick two-dimensional (2D) crystals[8-10] but also to create a vertical stack of them in any combination desired. The list of vdWs 2D crystals and the complexity of heterostructures[11] is growing by the day and so is their technological applications[12]. In device fabrication the interface between layered materials[13, 14] and the substrate (typically a non-vdWs material) is a complex region where vdWs and other related interactions[13, 15-19] can significantly influence the electrical, mechanical, and optical properties of layered materials. As such, the majority of work exploring these interactions has focused on characterization of the effects on electrical and optical properties[18]; however, interactions with the substrate can also impact the stability of layer materials. In particular the stability of air-sensitive vdWs materials such as phosphorus allotropes[20-24] can depend on interface forces that influence their interaction with their environment (water and oxygen in ambient). In addition, substrate hydrophilicity[25], and the extent of layered material/substrate contact are influenced by the interface vdWs forces and require careful investigation. It is then necessary to tune the interface vdWs forces and interface structure by selecting the right substrate[26] not only to optimize the physical properties of vdWs materials, but also to protect these materials from degradation preserving desired performance properties.

Violet phosphorus (VP) is an exciting recently rediscovered layered semiconductor with large tunable direct band gap up to 2.5 eV, intrinsic p-type conductivity, and high mobility[27, 28]. As such, it shows great promise for transistor, emitter, and photodetector applications, competitive with the other intrinsic p-type layered semiconductors. Many of which, BP[20, 21, 23, 29, 30], GeAs, and GaSe[31] are quite unstable under ambient conditions. For some time, BP was regarded as the most stable of the phosphorus allotropes, but recently VP was found to be more resistant to oxidation and have a higher thermal decomposition temperature making VP more suitable for ambient applications[32].

In this work, we investigate the interaction of exfoliable VP with various substrates such as $SiO_2$, mica, Si, and Au under long term exposure to air. We use a combination of infrared (IR) nanoimaging and theoretical modeling to study how the interface vdWs force between VP and the substrate affects its interactions with its surroundings, crucial for its stability. We find that in addition to substrate hydrophobicity and the extent of contact between VP and the substrate can also influence the interface vdWs forces. VP is found to be most stable when exfoliated on $SiO_2$ substrate, followed by mica, Au and is least stable on Si substrate. The results of this study could guide the selection of substrates when preparing vdW materials and more generally highlight the key role of interface force effects in altering the physical properties of vdW materials.

## RESULTS AND DISCUSSION

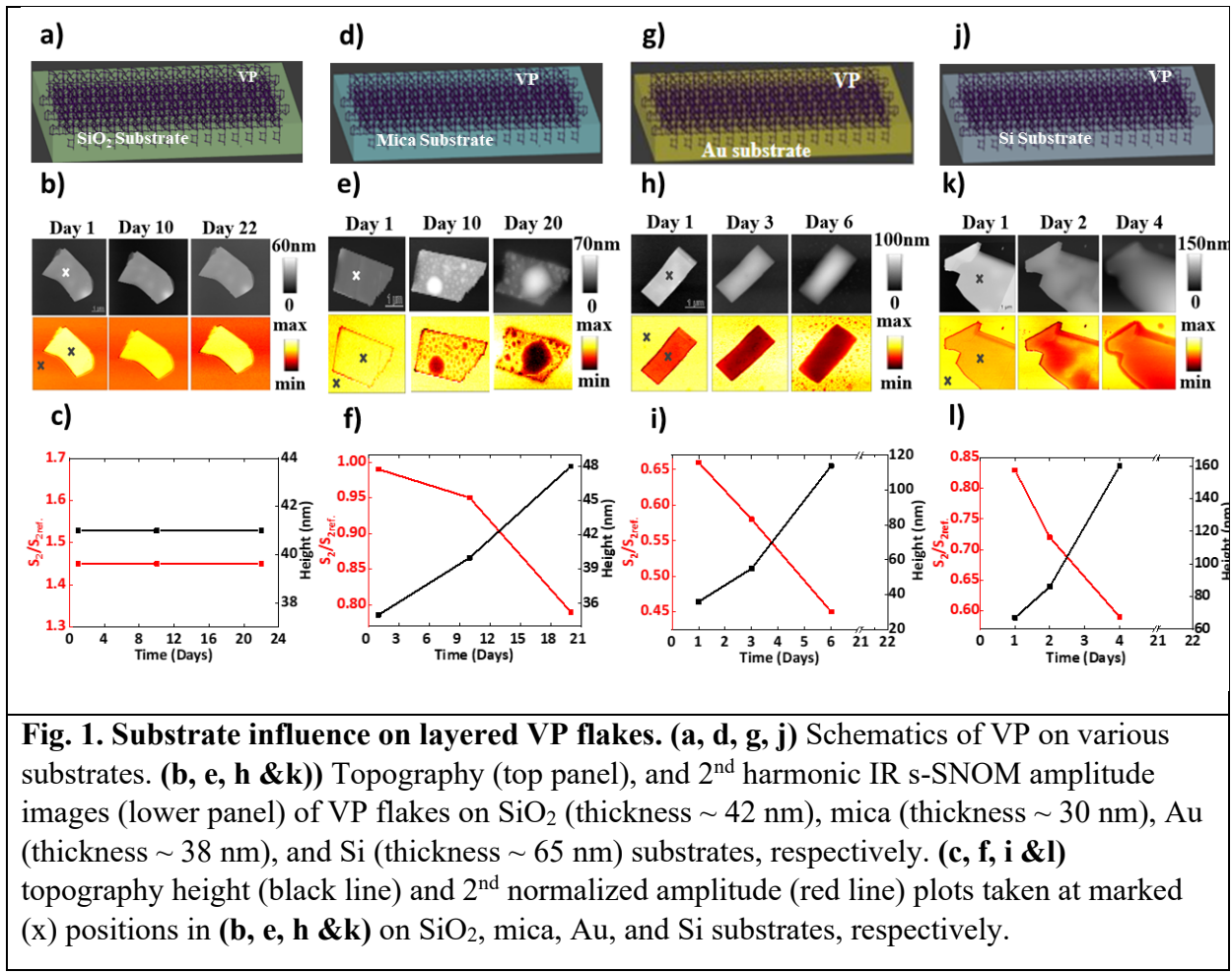

**Fig. 1. Substrate influence on layered VP flakes. (a, d, g, j)** Schematics of VP on various substrates. **(b, e, h &k))** Topography (top panel), and 2$^{nd}$ harmonic IR s-SNOM amplitude images (lower panel) of VP flakes on SiO$_2$ (thickness ~ 42 nm), mica (thickness ~ 30 nm), Au (thickness ~ 38 nm), and Si (thickness ~ 65 nm) substrates, respectively. **(c, f, i &l)** topography height (black line) and 2$^{nd}$ normalized amplitude (red line) plots taken at marked (x) positions in **(b, e, h &k)** on SiO$_2$, mica, Au, and Si substrates, respectively.

Interaction at the interface between vdWs materials and the substrate[33] on which they are exfoliated play a significant role modifying the intrinsic physical properties of layered materials. We investigate the interaction of exfoliated VP flakes (acquired from HQ Graphene) with several substrates using IR nanoimaging (see Methods). Figure 1 shows the experimental topographic and near-field amplitude measurements of exfoliated VP flakes over an extended time on four different substrates namely SiO$_2$, Mica, Au, and Si. The thickness of the flakes on different substrates were kept to ~30-65 nm range and all flakes on each substrate were measured consecutively using s-SNOM keeping all experimental conditions identical over several days. The four substrates were identically prepared and VP flakes of comparable thickness were exfoliated on them. Figure 1a shows schematic of exfoliated VP flake on SiO$_2$ substrate and Fig. 1b representative topography (black and white images) and near-field amplitude (red and yellow images) of a 42 nm thick VP flake on SiO$_2$ substrate imaged over several days. Both the topography and near-field amplitude images show no change for over 22 days, which is also shown in topography height (black line) and 2$^{nd}$ normalized amplitude (red line) plots taken at marked (x) positions in Fig. 1b. Each normalized point on the near-field amplitude line profile plot was acquired by taking signal at the locations marked x on the amplitude image normalized by a reference signal on the substrate. When we perform similar experiments on a 34 nm thick VP flake exfoliated on a mica substrate at the same time period and experimental conditions, we observe clear oxidation of VP as indicated by the change in the topography and the near-field amplitude images (Fig. 1e). The thickness of the VP flake increases by over 10 nm while the normalized amplitude values show a decreasing trend which confirms the oxidation of the VP flake. When VP (thickness ~38 nm) is exfoliated on an Au substrate, the flake completely degrades in 6 days resulting in ~80 nm topographic height increases and a corresponding normalized near-field amplitude decrease. The oxidation time is the shortest when a Si substrate is used for a ~65 nm VP flake. The sample completely oxidizes in 4 days and the topography height increases by ~100 nm simultaneously with a steep decrease in the normalized amplitude signal. Both an increase in the topographic height and decrease in time dependent near-field normalized amplitude signal are hallmarks of degradation as previously reported[20-23, 34]. The height changes are induced due to the formation of reaction species on the surface and bulk of oxidized VP where the adsorption of atmospheric water on $PO_x$[21, 35-38] results in these features and large height increases. The near-field amplitude decrease is because IR photons directly probe the changing permittivity of the sample over time as the sample degrades. Despite conducting a comparison study using data collected solely over a period of 22 days for $SiO_2$ substrate, the fact that VP flake did not degrade during this time frame does not lead to the conclusion that it will remain undegraded indefinitely; rather, it indicated that degradation did not occur within 22-day frame but is likely to occur over an extended period of time. Based on these results in the experimental conditions considered here VP is stable for a longer time in air if it is prepared on a $SiO_2$ substrate, followed by mica, Au and is least stable on a Si substrate.

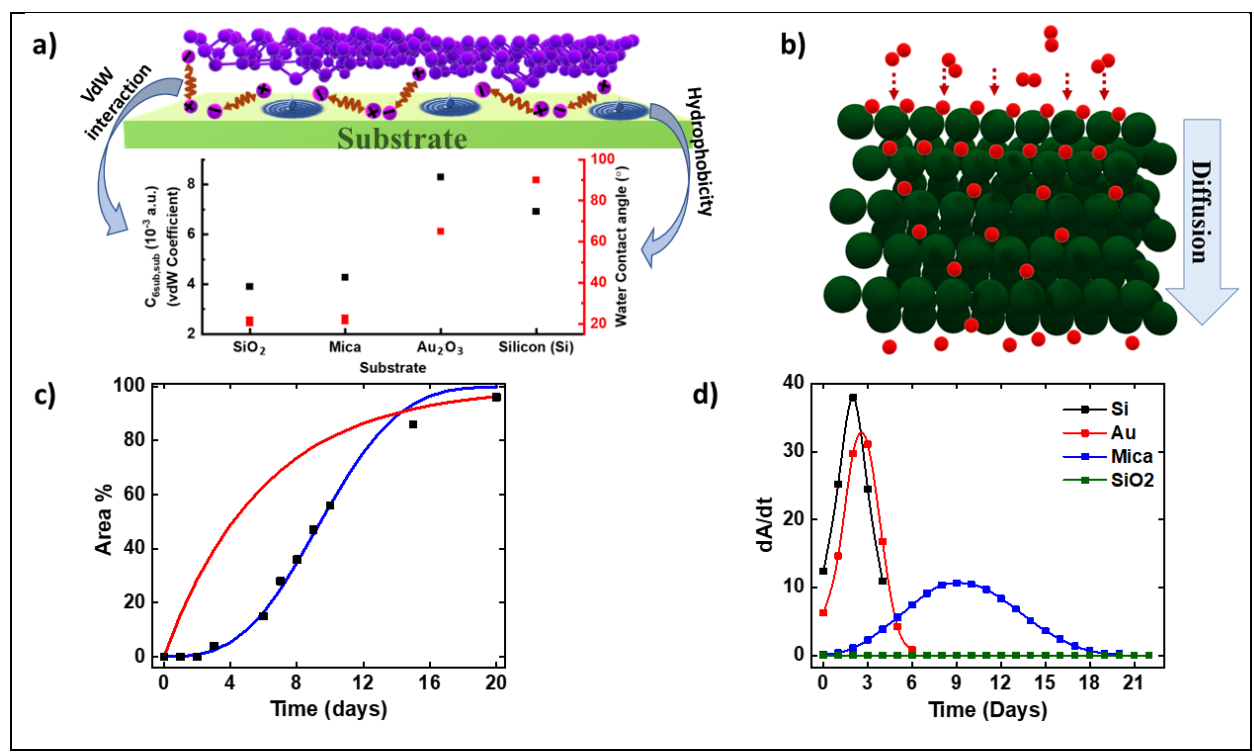

**Fig. 2 Effect of vdWs forces and hydrophilicity/hydrophobicity. a,** vdWs interaction of VP with a substrate and hydrophilicity/hydrophobicity of substrate with the violet phosphorous and, the graph showing values of vdW coefficient and water contact angles for $SiO_2$, Mica, Gold (Au), and Silicon (Si) substrates respectively. **b,** Schematic of diffusion of oxygen and water through the crystal in the ambient. **c,** show the experimental (black squares) and theoretical fit (blue and red curves) of area degradation using forest fire model for VP on mica substrates. **d,** Variation of dA/dt with time for the different substrates.

In this work, we consider two characteristics of the substrate surface (shown in Fig. 2a): (i) strength of vdW interactions with exfoliated VP flakes and (ii) hydrophilicity. To quantify the strength of the vdW interaction, we constructed a simple theory based on the first principles vdW calculations (see below). For the hydrophilicity effect, we considered experimentally available water contact angle (WCA) for simplicity. Based on these two characteristics, we propose a mechanism to explain the substrate-dependent stability of the exfoliated VP flakes.

In our proposed mechanism, the vdW interactions between VP and substrate determine the porosity (and guest-molecule dwell time) of the exfoliated VP flake. More specifically, the weaker vdW interactions lead to a larger gap at the interface (higher porosity and smaller guest-molecule dwell time), thereby opening ventilation pathways for reactants ($H_2O$ and $O_2$)[39] and the reactive intermediates to diffuse away from VP and slow down degradation. To quantify the relative strength of vdW interactions across the four substrates considered, we employed a simple vdW coefficient combining rule

$$C_{6VP,Sub} \approx \sqrt{C_{6VP,VP} C_{6Sub,Sub}},$$

in which $C_{6VP,Sub}$ is the vdW coefficient between VP and substrate, and $C_{6VP,VP}$ and $C_{6Sub,Sub}$ are self vdW coefficients for VP and the given substrate, respectively. We note that this geometric-average approach is one of the simplest combining rules for homo-species vdW coefficients (which assumes the average energies of the involved species are similar) and can be further refined with atomic polarizabilities. In doing so, the strength of the VP--substrate vdW interactions can be ranked for each substrate by simply comparing $C_{6Sub,Sub}$. However, such ranking requires normalized $C_{6Sub,Sub}$ values since the substrates used in this work have different unit cells. More specifically, we chose to consider each material as a continuum media and normalize their effective self vdW coefficient as

$$\bar{C}_{6Sub,Sub} = \frac{C_{6Sub,Sub}}{V^2},$$

in which V is the volume of the unit cell. In doing so, we compute $C_{6Sub,Sub}$ based on the atomistic structures in the unit cell via

$$C_{6Sub,Sub} = \sum_{A,B \in Sub} C_{6A,B},$$

in which $C_{6A,B}$ is the vdW coefficient between atoms A and B within the substrate unit cell. $C_{6AB}$ is computed directly using the Tkatchenko and Scheffler approach[40]. To treat ionic and covalent substrates on equal footing, we followed Hermann and Tkatchenko and modified with Vydrov and Van Voorhis VV10 polarizability[41] for scaling the free-atom reference to treat ionic and covalent substrates on equal footing[41]. All computations were performed using the FHI-AIMS package using the tight basis with k-point sampling sufficient to converge the self-consistent evaluation of $C_{6AB}$. As shown in Fig 2a, the computed $\bar{C}_{6Sub,Sub}$ [41] agrees quite well with our experimental observation, except that the order of Si and Au is inverted. We think this difference might originate from the lack of specific surface information for Au (e.g., oxidation state and surface termination); for instance, $\bar{C}_{6Sub,Sub}$ of Au changes quite substantially from $10.5 * 10^{-3}$ au to $8.3 * 10^{-3}$ au upon oxidation to $Au_2O_3$. Nevertheless, our theoretical modeling (in conjunction with the above-mentioned hypothesis) seems to provide a consistent explanation for the experimentally observed substrate-dependent VP degradation.

Second, we consider the effect of substrate hydrophilicity on the stability of VP and its correlation with the effect of vdWs forces between VP and substrates discussed above. The right side of the y-axis and the red square points in Fig. 2a show the water contact angle (WCA) values reported in literature for the different substrates used in our experiment (x-axis)[42-47]. The plot suggests that the degradation rate of VP flakes increases with the increasing hydrophobicity of the substrate showing a similar trend to that of vdW coefficients plotted for the different substrates. Silicon is the most hydrophobic with a WCA of 90°[42] and exfoliated VP flakes on it degrade at the fastest rate (in just 4 days). On the other hand, Mica[43] and $SiO_2$[44-47] substrates are hydrophilic with contact angles of 22° and 21° respectively, so exfoliated VP flakes on these substrates degrade at the slowest rate. Although error bars have been included in the graph to represent the contact angle values for $SiO_2$ and Mica, their size is so minute that they are not discernible in the visual representation. We have added the VP flakes on gold substrates degrade at an intermediate rate which has a contact angle of 65° (hydrophilicity in between). Since hydrophobic surfaces repel water, the water will tend to move in order to reduce the surface energy of the system forming large droplets or collecting at a more hydrophilic defect site. As VP oxidizes and forms hydrophilic $PO_x$ species [14], moisture could diffuse and collect at a VP flake contributing to the complex water/oxygen/light oxidation of chemistry of VP.

Next, we introduced a variant of the forest fire model earlier[20] to get qualitative insight on how surface morphology of vdW materials evolved as a function of time due to chemical oxidation. The model is based on a simple probabilistic description of the sample surface which is divided into N × N square elements with each element either in degraded or in undegraded state. The function $P_n = 1 - e^{-\Delta t \cdot \eta^{(n)}}$ gives the degradation probability after $\Delta t$ time interval, where degradation probability per unit time is given by $\eta^{(n)}$ with n degraded neighbors ($0 \leq n \leq 8$) is merely a fitting parameter. $\Delta \eta$ is the fixed increase in each degraded neighbor and can be written as $\eta^{(n)} = \eta^{(0)} + n\Delta\eta$, where $\eta^{(0)}$ is the degradation probability per unit time if all the neighbors are undegraded. The model provides several useful qualitative insights into the surface morphological modification process. As an example, in Fig. 2c we show experimental degradation area percentage data in black squares as a function of time when a mica substrate is used (we note that the trend is similar on all other substrates). The blue line shows the theoretical model fitting the percentage degradation area including the neighboring interaction (non-zero $\eta^{(0)}$). We found a good fit with the experimental results using $\eta^{(0)} = 0.015$ and $\Delta\eta = 1.65d^{-1}$. The red line shows a fitting, with all parameters kept the same, but without neighboring interactions included ($\eta^{(0)} = 0$). As shown in Fig. 2c the experimental data fits better when we consider non-zero neighboring interaction. This can be explained by invoking a diffusion model (schematics shown in Fig. 2b). As oxygen diffuses into the VP layer it collides with phosphorous atoms and if the phosphorus atoms are still unreacted and available it has a high probability of interacting. If on the other hand the phosphorous atoms are all reacted, the oxygen atoms will simply bounce off and move to the next neighboring phosphorus. In this way a "degraded/reacted" phosphorus makes it more probable for an undegraded region of the crystal to degrade. To show the difference in the degradation rate of VP on different substrates used, we find that it is more instructive to plot the slope of the degraded area (A) as a function of time (dA/dt vs. t) as shown in Fig. 2d. The plots show a Gaussian distribution of the slopes for Si, Mica and Au substrates and the substrate with the largest maximum slope (dA/dt) values will always degrade VP the fastest. Since the VP flake does not degrade on $SiO_2$ substrate in the period shown in the plots, the slope is zero. VP on Mica shows the slowest degradation next to $SiO_2$ substrate.

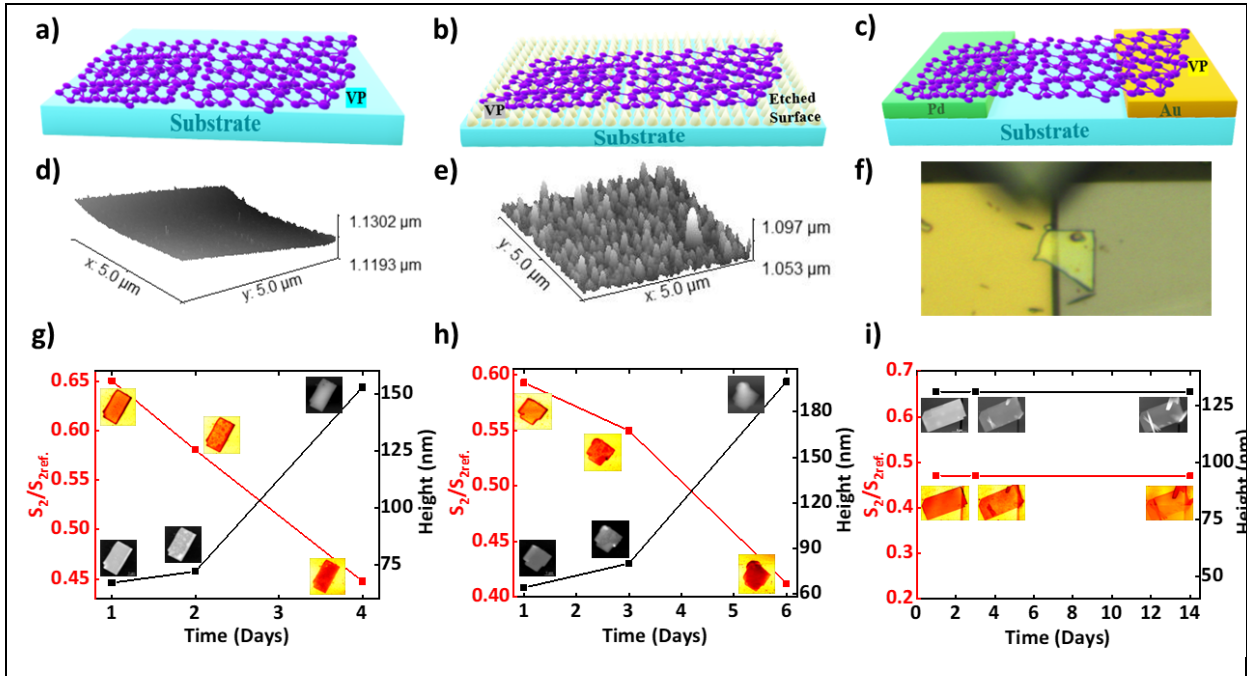

**Fig. 3 Modifying vdWs Forces using contact area on Degradation.** Schematics of VP flake **a,** on Si substrate etched less than one second. **b,** on a six hundred-second etched Si substrate. **c,** VP flake hanging on the two electrodes one having Pd and another having Au. **d-e,** 3D topographic images of <1 sec and 600 sec etched substrates, respectively. **f,** optical image of VP flake over the gap between the two electrodes. **g-i,** Variation of Height and normalized second harmonic amplitude values with the time along with the images of Topography and second harmonic amplitude images of <1 sec etched, 600 sec etched substrates and of Flake over the electrodes, respectively.

We further investigate the influence of vdWs force between exfoliated VP degradation by preparing VP flakes of similar thickness (~20 nm) on two Si substrates that are etched to produce different roughness as a way to reduce VP interaction with the substrate Fig. 3 d &e). Further, we compare the degradation of these samples to a suspended VP flake surrounded by air which does not interact with substrate (Fig. 3f). To produce Si substrates with different surface roughness (Fig. 3e) we etch the surface using a buffered oxide etch solution comprises 40% $NH_4F$ in water and 49% HF in water in a ratio of 6:1 for either less than one second or six hundred seconds, resulting in roughness values of 3 and 22nm, before we exfoliate VP. Due to the mechanical rigidness of the ~20 nm thick flakes used they do not conform to the rough etched surfaces as observed by the smooth VP topography. Based on the topography and normalized second harmonic amplitude graph (Fig. 3g) from VP on the smoother (3 nm roughness) Si substrate rapidly increased from 60 to 160 nm, and normalized amplitudes decreased linearly over 4 days. Figure 3e shows data from the VP flake on the rough (22 nm roughness) Si substrate. This sample has completely degraded in six days; the height has increased more than three times, as seen in the plot Fig. 3h), and the normalized amplitude value has decreased from 0.60 to 0.40. In the third sample VP is suspended between two electrodes and does not interact with any substrate (Fig. 3c&f). The height and normalized near-field amplitude plots show that VP flake on the smooth Si substrate with the largest vdWs interaction force degrades the fastest and the suspended flake with the least substrate interaction experiences the slowest degradation. The flake on the rough Si substrate shows a degradation time between these two cases. These controlled experiments clearly show that the dependence of degradation phenomena of vdWs materials on the vdWs force between the samples and substrate, an effect that needs also be considered in device and sample manufacturing.

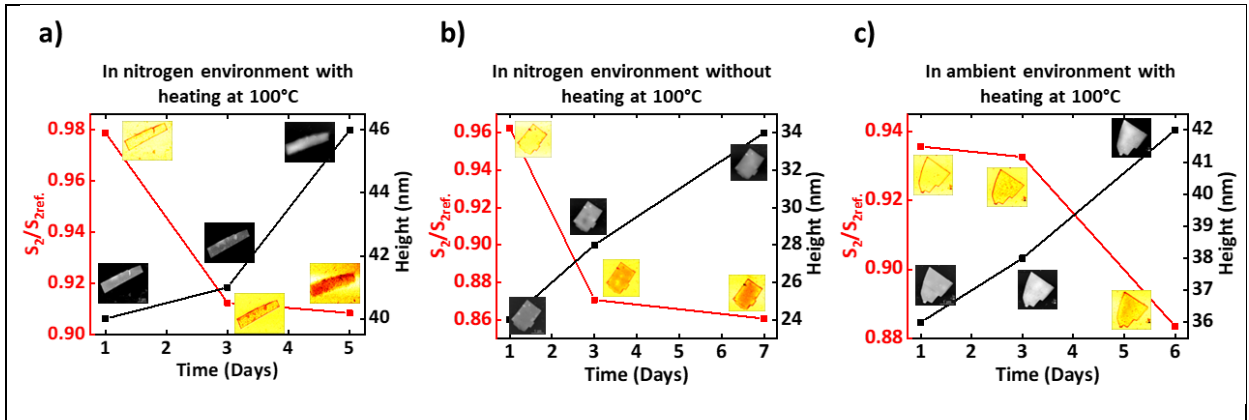

**Fig. 4 Effect of Oxygenated and deoxygenated environment on stability of VP exfoliated on Mica substrate.** Variation of the normalized second harmonic amplitude and height with the time (in days) with topography and second harmonic amplitude images taken with IR s-SNOM at a frequency of 952 $cm^{-1}$ in the inset of VP flakes exfoliated **a,** in nitrogen glove box without heating the substrate at 100°C before exfoliation the VP flake. **b,** in a nitrogen glove box with heating the substrate before exfoliation. **c,** in air with heating the substrate at 100°C before exfoliation of the flake.

Finally, we compare degradation of VP prepared on mica substrates exfoliated in air and nitrogen environments as well as the effect of heating the mica substrate. The degradation of VP on Mica substrate prepared under ambient environment is shown above in Fig. 1 e & f. We now used a nitrogen glove box (SI Fig. 3) and prepared a VP sample under deoxygenated low humidity conditions. We first exfoliated VP in a nitrogen environment without heating the mica substrate (Fig. 4a). This results in degraded VP within 7 days of exfoliation. We then heat the mica substrate to 100°C for 2 to 3 minutes and bring it to room temperature while in the glove box and then exfoliate VP (Fig. 4b), Although the flake was prepared in a nitrogen glove box, it still degraded completely within five days. Finally, after heating the mica substrate to 100°C in ambient conditions and then cooling it to room temperature (Fig. 4c), we exfoliated VP flake on it. In this case, the sample degraded completely in 6 days. We found that whenever we heated the substrate or put it in a deoxygenated environment, the pace of degradation increases. From these results, we speculate that when we heat the mica substrate prior to VP exfoliation water is driven off. Due to mica's hydrophilic nature, it attracts water from the environment as we measure the sample in ambient conditions. As a result, our results suggest that this process promotes oxidation of VP under the ambient environment due to heating of the substrate.

**CONCLUSION**

We investigated the effect of vdWs forces and hydrophobicity of various substrates on the chemical stability of exfoliated VP flakes using the nanoscale IR imaging techniques. VP shows strong stability preference when exfoliated on $SiO_2$ substrate as compared to other substrates such as Mica, Au, and Si. We found that the strength of vdWs forces plays a major role in the stability of VP. Additionally, we examine the effects of oxygenated and deoxygenated environments on VP degradation. Our results indicate that unetched substrates with stronger vdWs interactions, could lead to faster degradation of VP. We note that interaction between vdWs materials and their substrates is very complex, involving multi-electron interactions, electronic energy level alignment, and chemical complexities, among other factors, which are not explored in the current study. Consequently, our research could set the stage for more detailed investigations into interface physics.

**METHODS**

**IR nanoimaging**

Topographic and IR nanoimaging of VP flakes were performed using a commercial IR scattering-type scanning near-field microscope (s-SNOM) from neaspec co. s-SNOM is based on a tapping mode AFM equipped with a cantilevered metal-coated tip. The tip oscillates at a resonance frequency $\Omega \sim 270kHz$ and a tapping amplitude of ~100nm. An IR quantum cascade laser ($\lambda$=10.5μm) is focussed by a parabolic mirror on to the metalized tip and interacts with the sample. The scattered light from the tip-sample junction is demodulated at the second harmonics of the tip resonance frequency ($2\Omega$) and detected by phasemodulation (pseudo-heterodyne) interferometry, resulting simultaneous topography, optical amplitude, and phase images. s-SNOM provides material constrast iamging with a spatial resolution down to ~10 nm independent of the incident wavelength used from visible to terahertz frequency range. [48-50].

**Material Synthesis**

To synthesize VP, the process involved loading a quartz crucible into the CVD chamber. The crucible was filled with 0.1g of $SnI_4$ and 0.1g of Sn metal. A sapphire wafer was placed on top of the crucible. The chamber was then evacuated and filled with a mixture of $N_2$ and $H_2$ in a ratio of 95% to 5%, reaching a pressure of 700 Torr. Subsequently, the reactor was heated to 600°C, and the phosphorus source TBP was introduced at a flow rate of 40 sccm along with 160 sccm of $N_2$ and $H_2$ for a duration of 30 minutes. While maintaining the flow of TBP and $N_2$:$H_2$, the reactor gradually cooled to 400°C. Once the desired temperature was reached, the TBP flow was stopped, and the reactor was further cooled to room temperature. Finally, tiny VP flakes that had formed on the crucible and sapphire wafer edges were transferred to different substrates for characterization using a polydimethylsiloxane (PDMS) stamp.

**FUNDING**

Support for this work is provided by the Air Force Office of Scientific Research (AFOSR) grant number FA9550-19-0252.

**REFERENCES**

**Uncategorized References**

(1) Feynman, R. There's plenty of room at the bottom. In *Feynman and computation*, CRC Press, 2018; pp 63-76.
(2) Tkatchenko, A. Current understanding of van der Waals effects in realistic materials. *Advanced Functional Materials* **2015**, *25* (13), 2054-2061.
(3) Parsegian, V. A. *Van der Waals forces: a handbook for biologists, chemists, engineers, and physicists*; Cambridge university press, 2005.
(4) French, R. H.; Parsegian, V. A.; Podgornik, R.; Rajter, R. F.; Jagota, A.; Luo, J.; Asthagiri, D.; Chaudhury, M. K.; Chiang, Y.-m.; Granick, S. Long range interactions in nanoscale science. *Reviews of Modern Physics* **2010**, *82* (2), 1887.
(5) Langbein, D. Theory of van der Waals Attraction. *Springer tracts in modern physics* **1974**, 1-139.
(6) Dzyaloshinskii, I. E.; Lifshitz, E. M.; Pitaevskii, L. P. The general theory of van der Waals forces. *Advances in Physics* **1961**, *10* (38), 165-209.
(7) Hermann, J.; DiStasio Jr, R. A.; Tkatchenko, A. First-principles models for van der Waals interactions in molecules and materials: Concepts, theory, and applications. *Chemical Reviews* **2017**, *117* (6), 4714-4758.
(8) Novoselov, K.; Mishchenko, o. A.; Carvalho, o. A.; Castro Neto, A. 2D materials and van der Waals heterostructures. *Science* **2016**, *353* (6298), aac9439.
(9) Geim, A. K.; Novoselov, K. S. The rise of graphene. *Nature materials* **2007**, *6* (3), 183-191.
(10) Lu, Z.; Dunn, M. L. van der Waals adhesion of graphene membranes. *Journal of Applied Physics* **2010**, *107* (4), 044301.
(11) Geim, A. K.; Grigorieva, I. V. Van der Waals heterostructures. *Nature* **2013**, *499* (7459), 419-425.

(12) Lv, R.; Robinson, J. A.; Schaak, R. E.; Sun, D.; Sun, Y.; Mallouk, T. E.; Terrones, M. Transition metal dichalcogenides and beyond: synthesis, properties, and applications of single-and few-layer nanosheets. *Accounts of chemical research* **2015**, *48* (1), 56-64.
(13) Auerbach, S. M.; Carrado, K. A.; Dutta, P. K. *Handbook of layered materials*; CRC press, 2004.
(14) Duong, D. L.; Yun, S. J.; Lee, Y. H. van der Waals layered materials: opportunities and challenges. *Acs Nano* **2017**, *11* (12), 11803-11830.
(15) Rösner, M.; Şaşıoğlu, E.; Friedrich, C.; Blügel, S.; Wehling, T. Wannier function approach to realistic Coulomb interactions in layered materials and heterostructures. *Physical Review B* **2015**, *92* (8), 085102.
(16) Cygan, R. T.; Greathouse, J. A.; Heinz, H.; Kalinichev, A. G. Molecular models and simulations of layered materials. *Journal of Materials Chemistry* **2009**, *19* (17), 2470-2481.
(17) Marom, N.; Bernstein, J.; Garel, J.; Tkatchenko, A.; Joselevich, E.; Kronik, L.; Hod, O. Stacking and registry effects in layered materials: the case of hexagonal boron nitride. *Phys Rev Lett* **2010**, *105* (4), 046801.
(18) Li, M.-Y.; Chen, C.-H.; Shi, Y.; Li, L.-J. Heterostructures based on two-dimensional layered materials and their potential applications. *Materials Today* **2016**, *19* (6), 322-335.
(19) Zhou, X.; Hu, X.; Yu, J.; Liu, S.; Shu, Z.; Zhang, Q.; Li, H.; Ma, Y.; Xu, H.; Zhai, T. 2D layered material-based van der Waals heterostructures for optoelectronics. *Advanced Functional Materials* **2018**, *28* (14), 1706587.
(20) Abate, Y.; Gamage, S.; Li, Z.; Babicheva, V.; Javani, M. H.; Wang, H.; Cronin, S. B.; Stockman, M. I. Nanoscopy reveals surface-metallic black phosphorus. *Light-Sci Appl* **2016**, *5*. DOI: ARTN e16162

10.1038/lsa.2016.162.
(21) Fali, A.; Snure, M.; Abate, Y. Violet phosphorus surface chemical degradation in comparison to black phosphorus. *Appl Phys Lett* **2021**, *118* (16). DOI: Artn 163105

10.1063/5.0045090.
(22) Gamage, S.; Fali, A.; Aghamiri, N.; Yang, L.; Ye, P. D.; Abate, Y. Reliable passivation of black phosphorus by thin hybrid coating. *Nanotechnology* **2017**, *28* (26). DOI: ARTN 265201

10.1088/1361-6528/aa7532.
(23) Gamage, S.; Li, Z.; Yakovlev, V. S.; Lewis, C.; Wang, H.; Cronin, S. B.; Abate, Y. Nanoscopy of Black Phosphorus Degradation. *Adv Mater Interfaces* **2016**, *3* (12). DOI: ARTN 1600121

10.1002/admi.201600121.
(24) Peng, L.; Abbasi, N.; Xiao, Y.; Xie, Z. J. Black Phosphorus: Degradation Mechanism, Passivation Method, and Application for In Situ Tissue Regeneration. *Adv Mater Interfaces* **2020**, *7* (23). DOI: ARTN 200153810.1002/admi.202001538.

(25) Wood, J. D.; Wells, S. A.; Jariwala, D.; Chen, K.-S.; Cho, E.; Sangwan, V. K.; Liu, X.; Lauhon, L. J.; Marks, T. J.; Hersam, M. C. Effective passivation of exfoliated black phosphorus transistors against ambient degradation. *Nano Lett* **2014**, *14* (12), 6964-6970.
(26) Mannix, A. J.; Kiraly, B.; Hersam, M. C.; Guisinger, N. P. Synthesis and chemistry of elemental 2D materials. *Nature Reviews Chemistry* **2017**, *1* (2), 0014.
(27) Schusteritsch, G.; Uhrin, M.; Pickard, C. J. Single-layered hittorf's phosphorus: a wide-bandgap high mobility 2D material. *Nano Lett* **2016**, *16* (5), 2975-2980.
(28) Ricciardulli, A. G.; Wang, Y.; Yang, S.; Samori, P. Two-Dimensional Violet Phosphorus: A p-Type Semiconductor for (Opto)electronics. *J Am Chem Soc* **2022**, *144* (8), 3660-3666. DOI: 10.1021/jacs.1c12931.
(29) Island, J. O.; Steele, G. A.; van der Zant, H. S.; Castellanos-Gomez, A. Environmental instability of few-layer black phosphorus. *2d Mater* **2015**, *2* (1), 011002.

(30) Abate, Y.; Akinwande, D.; Gamage, S.; Wang, H.; Snure, M.; Poudel, N.; Cronin, S. B. Recent progress on stability and passivation of black phosphorus. *Advanced Materials* **2018**, *30* (29), 1704749.
(31) Yagmurcukardes, M.; Senger, R. T.; Peeters, F. M.; Sahin, H. Mechanical properties of monolayer GaS and GaSe crystals. *Physical Review B* **2016**, *94* (24), 245407.
(32) Zhang, L.; Huang, H.; Zhang, B.; Gu, M.; Zhao, D.; Zhao, X.; Li, L.; Zhou, J.; Wu, K.; Cheng, Y. Structure and properties of violet phosphorus and its phosphorene exfoliation. *Angewandte Chemie* **2020**, *132* (3), 1090-1096.
(33) Bian, R.; Li, C.; Liu, Q.; Cao, G.; Fu, Q.; Meng, P.; Zhou, J.; Liu, F.; Liu, Z. Recent progress in the synthesis of novel two-dimensional van der Waals materials. *National Science Review* **2022**, *9* (5), nwab164.
(34) Abate, Y. Recent progress on stability and passivation of black phosphorus. *Abstr Pap Am Chem S* **2018**, *256*.
(35) Xue, J.; Wang, S.; Zhou, J.; Li, Q.; Zhou, Z.; Hui, Q.; Hu, Y.; Zhou, Z.; Feng, Z.; Yan, Q. Passivating violet phosphorus against ambient degradation for highly sensitive and long-term stable optoelectronic devices. *Appl Phys Lett* **2023**, *122* (18).
(36) Zhang, T.; Wan, Y.; Xie, H.; Mu, Y.; Du, P.; Wang, D.; Wu, X.; Ji, H.; Wan, L. Degradation chemistry and stabilization of exfoliated few-layer black phosphorus in water. *J Am Chem Soc* **2018**, *140* (24), 7561-7567.
(37) Luo, W.; Zemlyanov, D. Y.; Milligan, C. A.; Du, Y.; Yang, L.; Wu, Y.; Peide, D. Y. Surface chemistry of black phosphorus under a controlled oxidative environment. *Nanotechnology* **2016**, *27* (43), 434002.
(38) Wu, S.; He, F.; Xie, G.; Bian, Z.; Luo, J.; Wen, S. Black phosphorus: degradation favors lubrication. *Nano Lett* **2018**, *18* (9), 5618-5627.
(39) Huang, Y.; Qiao, J.; He, K.; Bliznakov, S.; Sutter, E.; Chen, X.; Luo, D.; Meng, F.; Su, D.; Decker, J. Interaction of black phosphorus with oxygen and water. *Chem Mater* **2016**, *28* (22), 8330-8339.
(40) Tkatchenko, A.; Scheffler, M. Accurate molecular van der Waals interactions from ground-state electron density and free-atom reference data. *Phys Rev Lett* **2009**, *102* (7), 073005.
(41) Hermann, J.; Tkatchenko, A. Density functional model for van der Waals interactions: Unifying many-body atomic approaches with nonlocal functionals. *Phys Rev Lett* **2020**, *124* (14), 146401.
(42) Bryk, P.; Korczeniewski, E.; Szymański, G. S.; Kowalczyk, P.; Terpiłowski, K.; Terzyk, A. P. What is the value of water contact angle on silicon? *Materials* **2020**, *13* (7), 1554.
(43) Rossi, E.; Phani, P. S.; Guillemet, R.; Cholet, J.; Jussey, D.; Oliver, W.; Sebastiani, M. A novel nanoindentation protocol to characterize surface free energy of superhydrophobic nanopatterned materials. *Journal of Materials Research* **2021**, *36* (11), 2357-2370.
(44) Martinez, N. *Wettability of silicon, silicon dioxide, and organosilicate glass*; University of North Texas, 2009.
(45) Stallard, C. P.; McDonnell, K. A.; Onayemi, O. D.; O'Gara, J. P.; Dowling, D. P. Evaluation of Protein Adsorption on Atmospheric Plasma Deposited Coatings Exhibiting Superhydrophilic to Superhydrophobic Properties. *Biointerphases* **2012**, *7* (1-4). DOI: ARTN 31

10.1007/s13758-012-0031-0.
(46) Frieser, R. Characterization of thermally grown SiO2 surfaces by contact angle measurements. *Journal of the Electrochemical society* **1974**, *121* (5), 669.
(47) Kim, E. K.; Kim, J. Y.; Kim, S. S. Significant change in water contact angle of electrospray-synthesized SiO2 films depending on their surface morphology. *Surface and interface analysis* **2013**, *45* (2), 656-660.
(48) Ogawa, Y.; Minami, F.; Abate, Y.; Leone, S. R. Nanometer-scale dielectric constant of Ge quantum dots using apertureless near-field scanning optical microscopy. *Appl Phys Lett* **2010**, *96* (6), 063107.

(49) Abate, Y.; Seidlitz, D.; Fali, A.; Gamage, S.; Babicheva, V.; Yakovlev, V. S.; Stockman, M. I.; Collazo, R.; Alden, D.; Dietz, N. Nanoscopy of Phase Separation in InxGa1-xN Alloys. *Acs Appl Mater Inter* **2016**, *8* (35), 23160-23166. DOI: 10.1021/acsami.6b06766.
(50) Aghamiri, N. A.; Hu, G.; Fali, A.; Zhang, Z.; Li, J.; Balendhran, S.; Walia, S.; Sriram, S.; Edgar, J. H.; Ramanathan, S. Reconfigurable hyperbolic polaritonics with correlated oxide metasurfaces. *Nature Communications* **2022**, *13* (1), 4511.